\documentclass{cernyrep} 
\usepackage{texnames}
\usepackage[T1]{fontenc}
\usepackage{subfig}
\begin{document}
\title{R\&D for Future 100 kton Scale Liquid Argon Detectors
\footnote{Contribution to the Workshop ``European Strategy for Future Neutrino Physics'', CERN, Oct. 2009, to appear in the Proceedings.}}
 
\author{A. Marchionni}

\institute{ETH Zurich, 101 Raemistrasse, CH-8092 Zurich, Switzerland}

\maketitle 

\begin{abstract}
Large liquid argon (LAr) detectors, up to 100 kton scale, are presently being considered for proton decay searches and neutrino astrophysics as well as far detectors for the next generation of long baseline neutrino oscillation experiments, aiming at neutrino mass hierarchy determination and CP violation searches in the leptonic sector. These detectors rely on the possibility of maintaining large LAr masses stably at cryogenic conditions with low thermal losses and of achieving long drifts of the ionization charge, so to minimize the number of readout channels per unit volume. Many R\&D initiatives are being undertaken throughout the world, following somewhat different concepts for the final detector design, but with many common basic R\&D issues. 
\end{abstract}
\section{Introduction}
\label{sec:intro}
A high granularity detector of large size, with a fine sampling down to a few percent of a radiation length, and with tracking and calorimetric capabilities, would be ideal to perform next generation neutrino physics, providing a clean identification of $\nu_{e}$ induced charged current interactions. Bubble chambers have clearly shown their potentiality as neutrino detectors, though limited in size, while larger size calorimetric detectors have been suffering from coarse granularity and limitations in the identification of the electromagnetic component in neutrino interactions. A clean measurement of electrons is becoming crucial in long baseline neutrino oscillation experiments driven by $\sim$MW proton beams for the determination of the $\theta_{13}$ mixing angle, the neutrino mass hierarchy and ultimately in the search for CP violation in the leptonic sector. Large liquid argon time projection chambers (LAr TPC), up to $\sim$100 kton size, have been proposed as far dectors in these experiments (\Refs~\cite{Rubbia:2004tz}--\cite{Badertscher:2008bp}). LAr TPCs, when compared to water Cerenkov detectors, allow lower momentum thresholds for the identification of heavier particles, notably protons, and are predicted to have higher electron identification efficiency with better rejection of the $\pi^{0}$ background.

A LAr detector of 100 kton size, if installed underground even at moderate depth, would extend the search for proton decay via modes favored by supersymmetric grand unified models (e.g. $p \longrightarrow K^{+} \bar{\nu}$) up to $\sim$10$^{35}$ years, having from 5 to 10 times the efficiency of water Cerenkov detectors for such decays~\cite{Bueno:2007um}. The synergy between precise detectors for long neutrino baseline experiments, proton decay and astrophysical neutrinos (see \Refs~\cite{GilBotella:2004bv,Cocco:2004ac}) is essential for a realistic proposal of a 100 kton scale LAr detector.
\begin{figure}[ht]
\centering\includegraphics[width=0.9\linewidth]{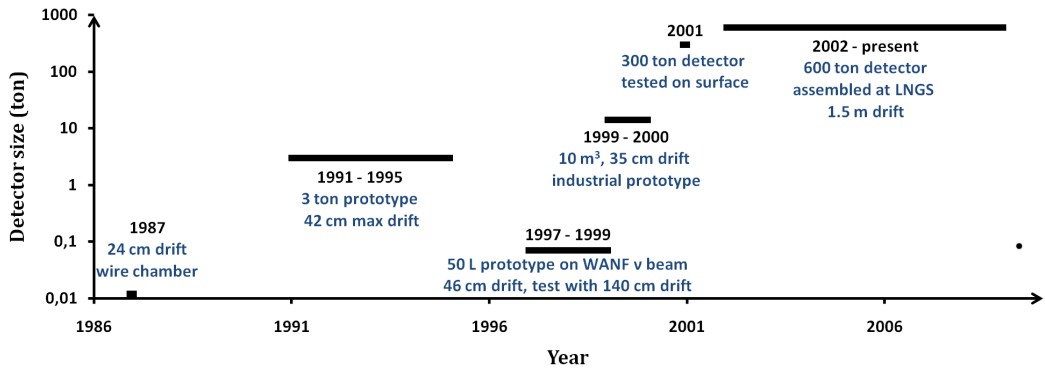}
\caption{The ICARUS R\&D steps towards kton size LAr detectors  }\label{fig:ICARUS_steps}
\end{figure} 

The LAr TPC technique, first proposed by C. Rubbia in 1977~\cite{CRubbia:1977}, has been developed in the last 20 years by the ICARUS collaboration (see \Fref{fig:ICARUS_steps}), culminating with a 300 ton detector (T300) successfully operated on surface ~\cite{Amerio:2004ze} and a 600 ton detector (T600) installed undeground in LNGS along the CNGS neutrino beam, now almost ready to be commissioned. 

\begin{figure}[b]
\centering
\begin{minipage}[t]{0.5\linewidth}
\centering\includegraphics[width=1.0\linewidth]{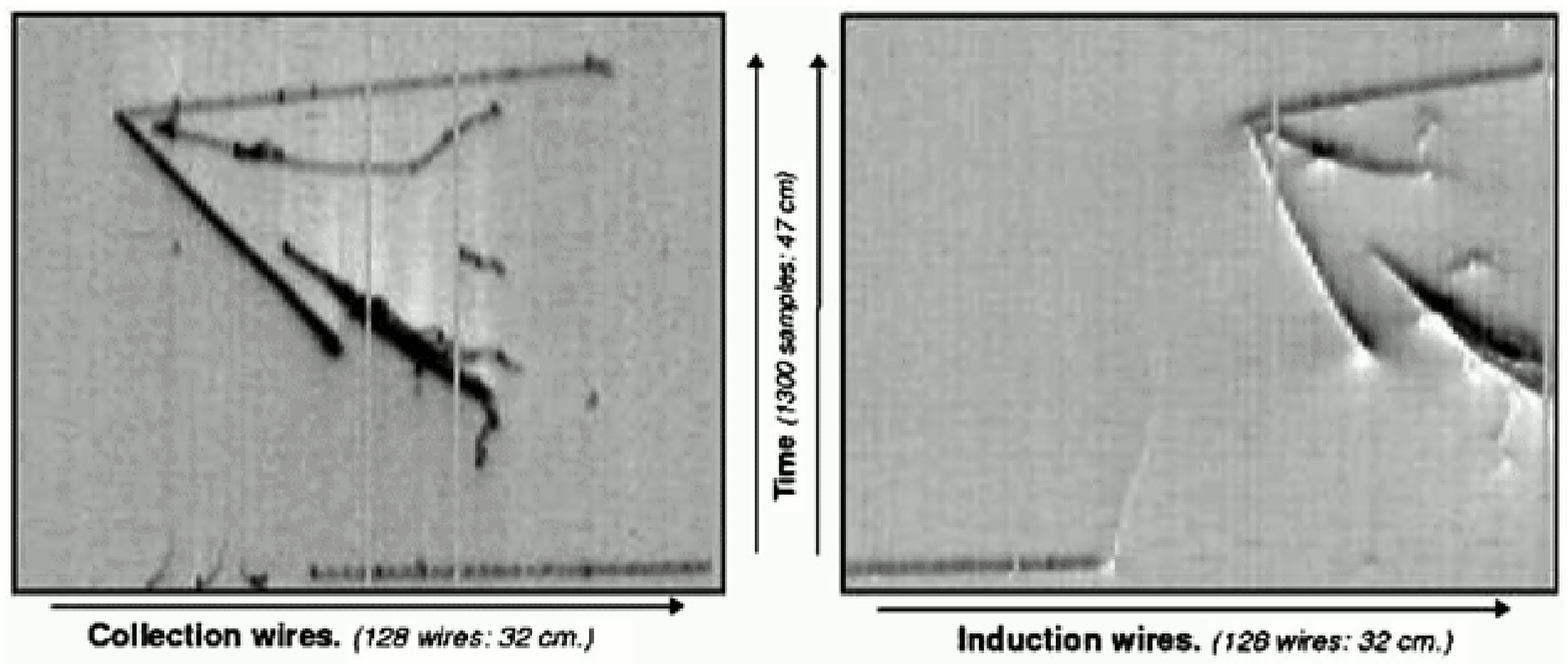}
\caption{An event recorded in the ICARUS 50 L chamber~\cite{Arneodo:2006ug}}\label{fig:ICARUS50lt}
\end{minipage}%
\hspace{0.5 cm}
\begin{minipage}[t]{0.40\linewidth}
\centering\includegraphics[width=0.65\linewidth]{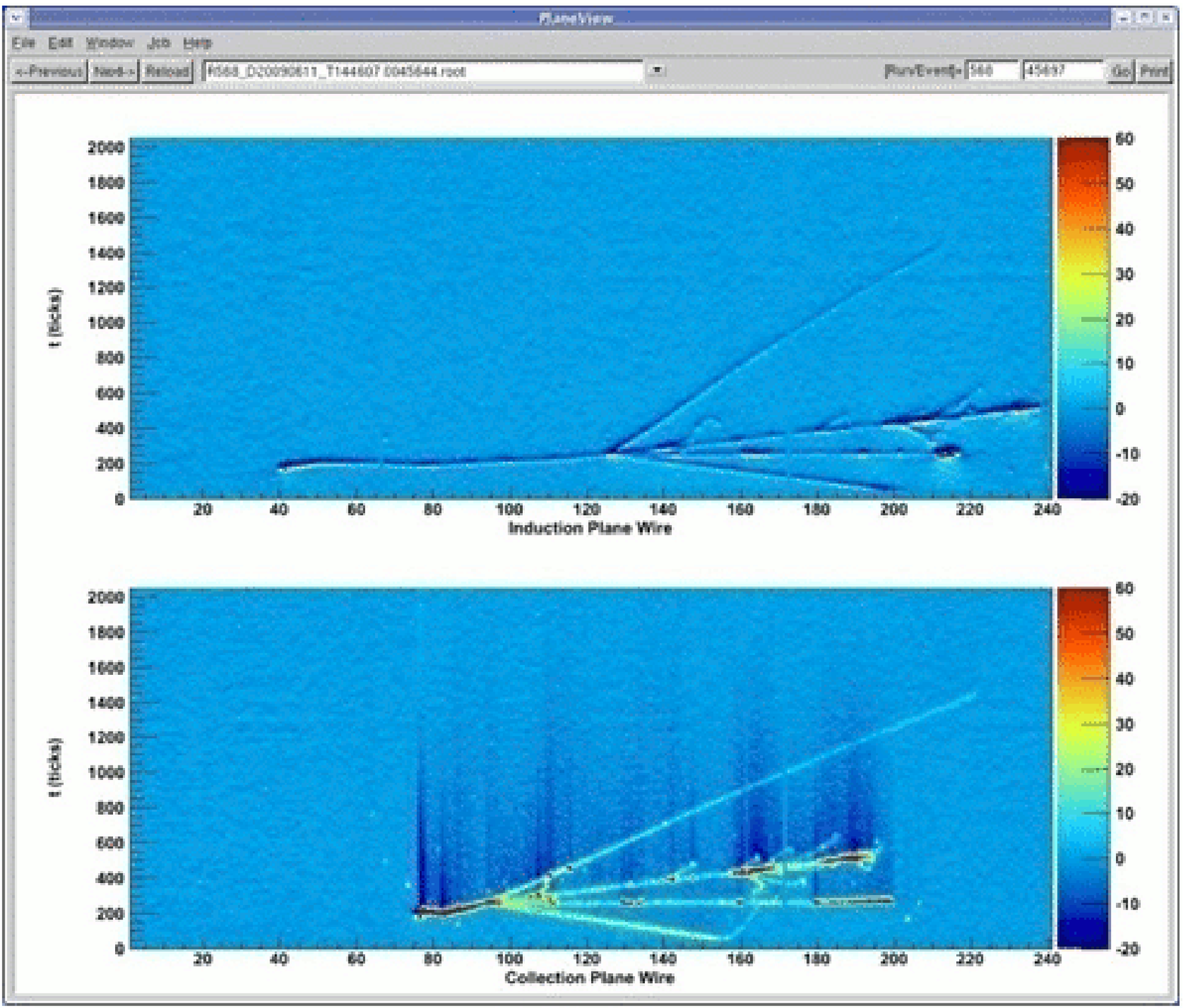}
\caption{An event in the ArgoNeut chamber on the NuMI beam at FNAL ~\cite{Fleming:2009}}\label{fig:ArgoNeut}
\end{minipage}
\end{figure}

The use of LAr TPCs as neutrino detectors was pioneered already ten years ago by the ICARUS collaboration with the exposure of a 50 liters LAr TPC ~\cite{Arneodo:2006ug} on the WANF neutrino beam (see \Fref{fig:ICARUS50lt}). Given the revived interest for the LAr technique, nowadays several groups have already placed ~\cite{Fleming:2009} or are planning to expose ~\cite{Maruyama:2009} somewhat bigger LAr detectors on neutrino beams, of 170 and 130 liters active volumes, respectively. About 20k neutrino interactions are expected by spring 2010 in the ArgoNeut detector installed on the NuMI neutrino beam at FNAL in front of the MINOS Near Detector (see \Fref{fig:ArgoNeut}).

Recently proposed LAr detectors require a step in size of about a factor 100 with respect to the ICARUS detector. Though challenging, this is not unrealistic over a period of $\sim$10 years and a well thought R\&D path. In Section~\ref{sec:techissue} we will present the main technical issues in scaling from an ICARUS-like detector to a $\sim$100 times larger device and in Section~\ref{sec:RandD} we will describe in some detail the ongoing R\&D work. Different design concepts have been proposed by different groups for 100 kton scale LAr TPCs. These are summarized in Section~\ref{sec:designs}, together with the proposed R\&D paths to get to the final detectors.
\section{Technical issues for large LAr TPCs}
\label{sec:techissue}
\begin{figure}[ht]
\centering\includegraphics[width=.53\linewidth]{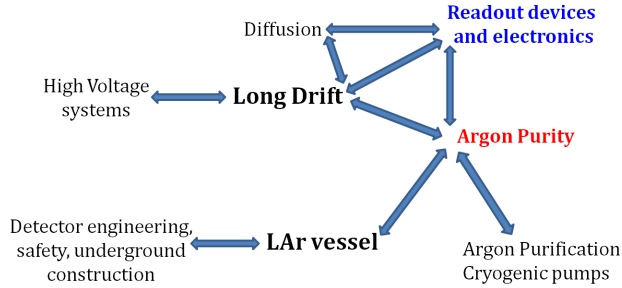}
\caption{Technical issues for large LAr TPCs}\label{fig:RandD}
\end{figure}

Fundamental physics questions would be answered by operating a 50--100 kton active volume LAr TPC detector. Extrapolation to $\sim$100 times larger size than the ICARUS T600 detector requires: 
\begin{enumerate}
\item longer drift path for the collection of the ionization charge, in order to reduce the number of readout channels per unit volume and the dead spaces introduced by the readout electrodes and cathode structures. As discussed in Section~\ref{sec:designs}, drift lengths ranging from a few to about 10 times the 1.5 m ICARUS T600 drift length have been proposed. Longer drift lengths can only be achieved if the ionization charge from minimum ionizing particles at the maximum drift distance is still detectable and suitable high voltage systems, that could produce the required drift field (0.5--1 kV/cm), are available.
\item larger cryogenic vessels, suitable for undeground construction and operation, with low enough thermal losses to afford the costs of a long term cryogenic operation and of sufficient tightness to keep electronegative impurities within a few tens of parts per trillion (ppt). Recirculation and purification systems in liquid and gas phases will be necessary to achieve and maintain the required LAr purity. Even in a modular approach, where the $\sim$100 kton LAr mass is achieved by multiple smaller detectors, LAr vessels of more than ten times the ICARUS size are required.  If on one hand this represents a challenge for the engineering design of the vessel, on the other hand a larger volume/surface ratio is an advantage for maintaining the detector in cryogenic conditions and for the purity of the LAr itself.
\end{enumerate}
The main technical issues are summarized in \Fref{fig:RandD}. An active R\&D program pursued by several groups in different regions of the world is already addressing these issues, as described in detail in the following Section.
\section{R\&D items}
\label{sec:RandD}
\subsection{Readout devices and electronics}
\label{sec:readout}
The choice of readout methods of the ionization charge released in LAr and the corresponding electronics plays an essential role in the design of a long drift LAr detector. For a given LAr purity, longer drifts are affordable only if the signal of a minimum ionizing particle, collected from the maximum drift distance, is sufficiently larger than the electronic noise (typically a signal/noise ratio of at least 10 is necessary), posing stringent requirements on the electronics and somehow favoring the use of readout devices where the released ionization charge is amplified. 

The standard method for reading out the ionization charge in LAr, decribed in \Refs~\cite{CRubbia:1977, Gatti:1979}, and pioneered by the ICARUS collaboration, makes use of three parallel wire planes, with wires at different angles. The first two wire planes are fully transparent to the drifting electrons and detect a signal induced by the moving electrons, while the third plane acts as a collecting electrode. By digitizing the signal as a function of the drift time, each wire plane provides a two dimensional view of the event, with one coordinate given by the wire number and the other by the drift time. Extrapolation of this technique to $\sim$100 kton LAr vessels is not straightforward. It would require very long wires if a single LAr vessel is considered, resulting, in addition to mechanical issues with long wires, in larger capacitances and increased noise, or, in order to limit the length of the wires, it would demand a modularized approach with several smaller independent vessels or a segmentation of the LAr volume with several independent readout wire planes. The increased capacitance for long wires has to be compensated by widening the wire pitch, as in ~\cite{Angeli:2009zza}, in order to regain a good signal/noise ratio.
\begin{figure}[ht]
\centering\includegraphics[width=.18\linewidth]{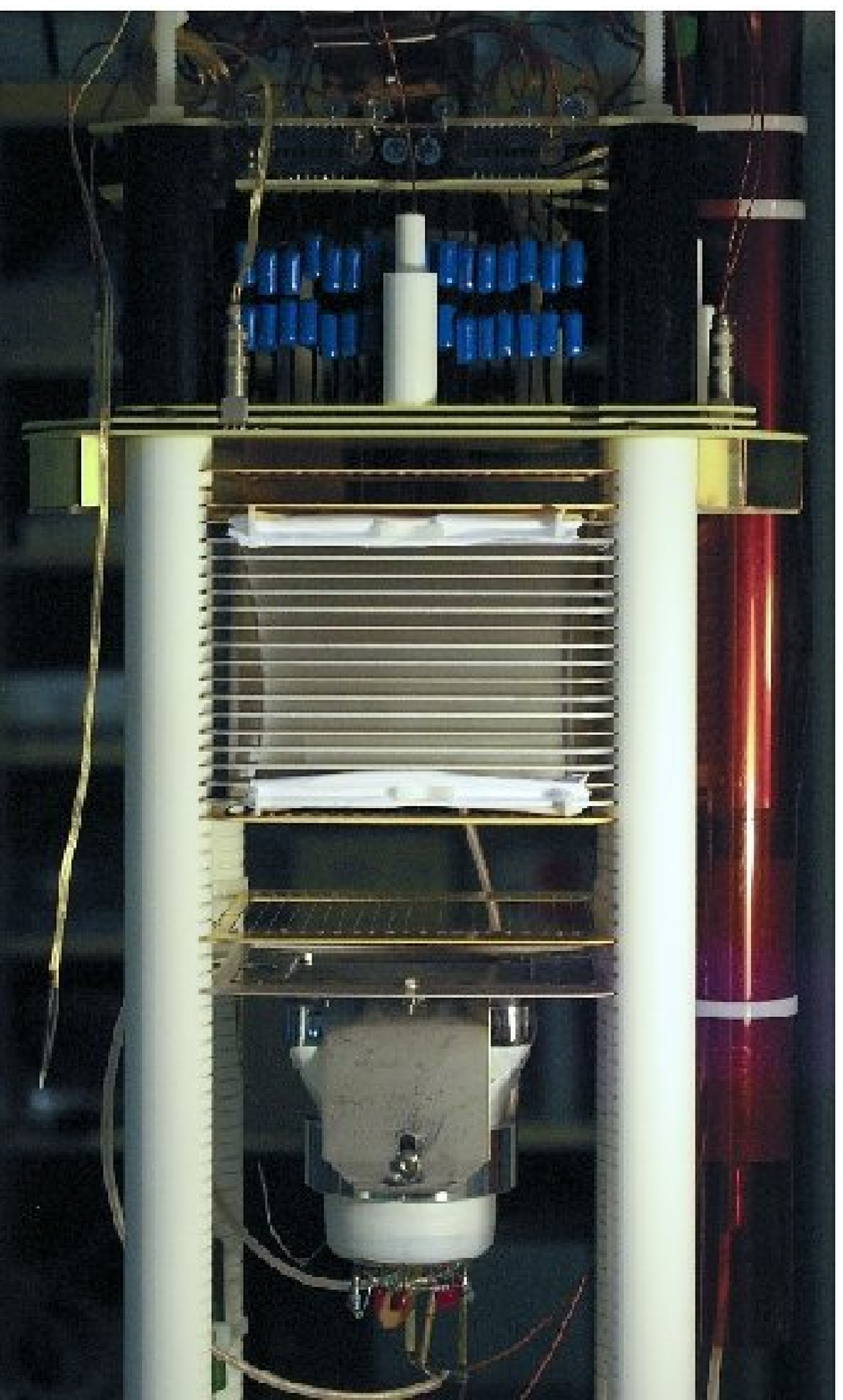}
\hspace{2 cm}
\centering\includegraphics[width=.35\linewidth]{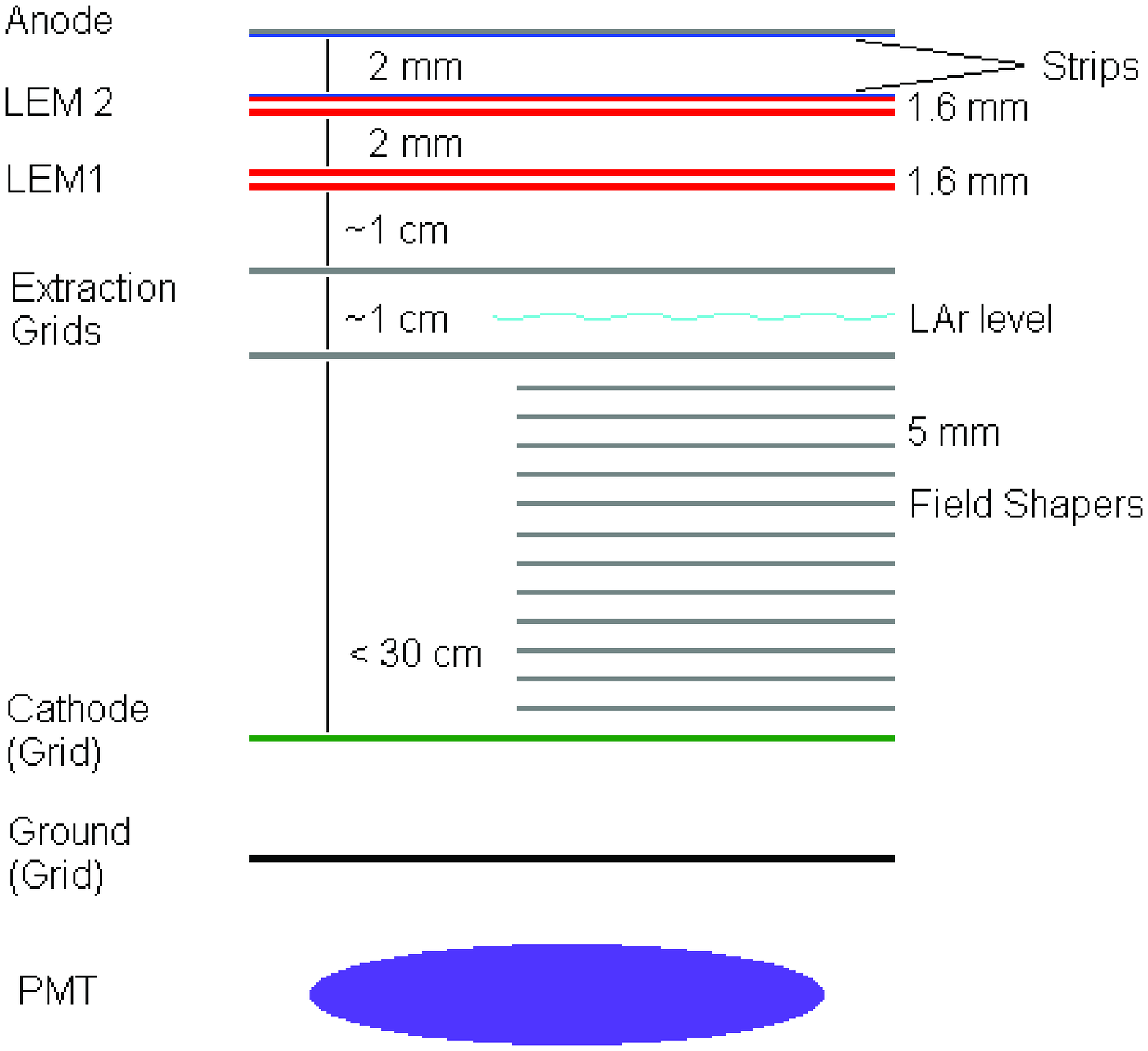}
\caption{Thick GEM LAr TPC: (left) photo of the test setup at CERN; (right) schematic of the chamber}\label{fig:THGEMTPC}
\end{figure}

The operation of a LAr TPC in double phase (liquid-vapor), as suggested in \cite{Rubbia:2004tz}, opens the possibility of amplification of the ionization charge in the pure argon vapor phase. A very active and promising R\&D program is being conducted by several groups on thick GEMs (THGEM, macroscopic hole multipliers manufactured with standard PCB techniques) as charge amplifying devices (see \Ref~\cite{Breskin:2008cb} for a recent review). They have been successfully operated in noble gases ~\cite{Alon:2008zz}, without any added quenching gas, and in double-phase argon detectors ~\cite{Polina_ETHthesis, Bondar:2008yw}. THGEMs represent a robust and economic way to realize large area detectors, suitable for cryogenic operation.

\begin{figure}[ht]
\subfloat[Crossing muon events: (top) electronic signals; (bottom) displays of drift time vs position]{
\label{fig:1mm_THGEM:a} 
\begin{minipage}[b]{0.60\linewidth}
\centering \includegraphics[width=0.47\linewidth]{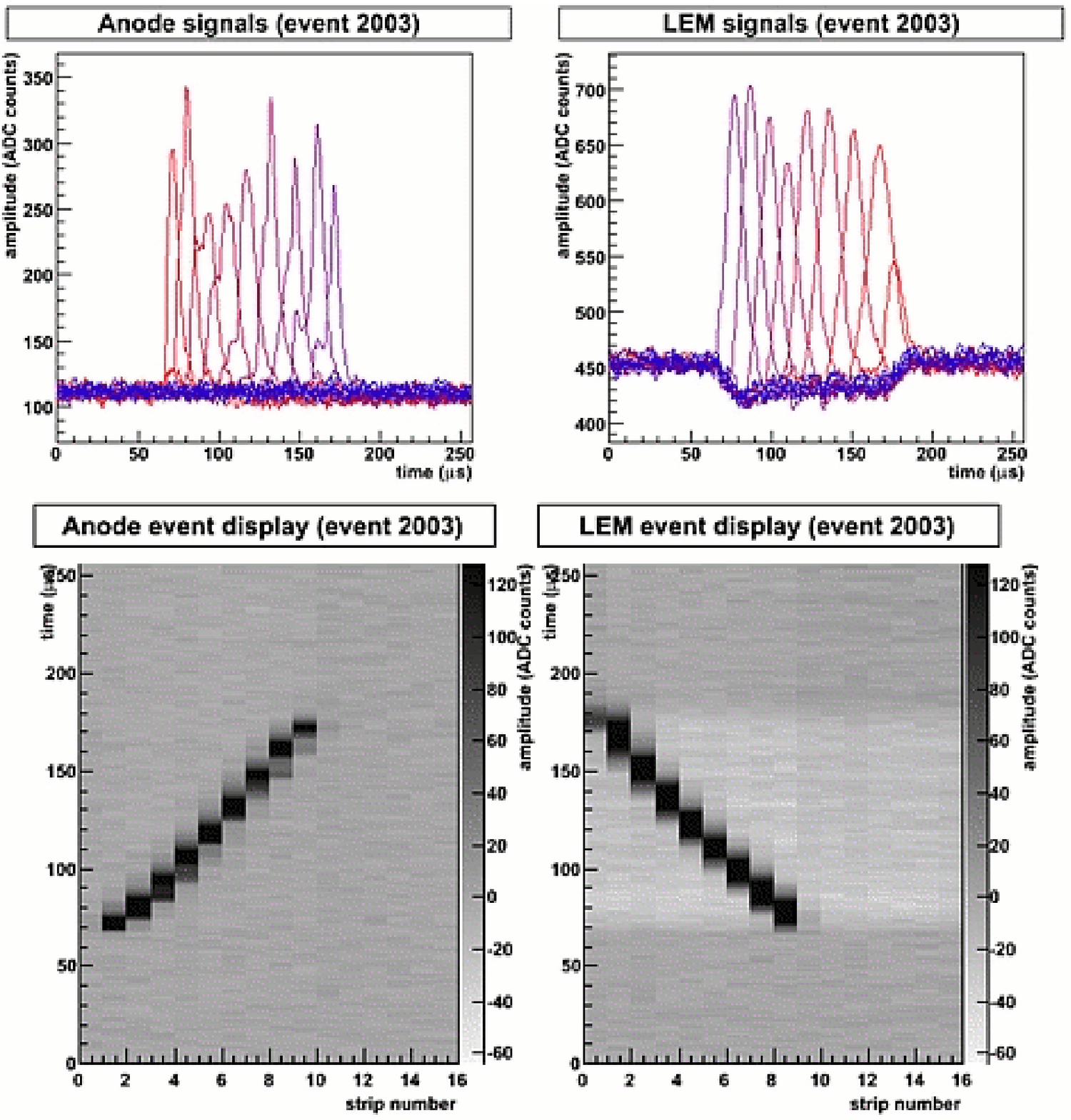}
\centering \includegraphics[width=0.45\linewidth]{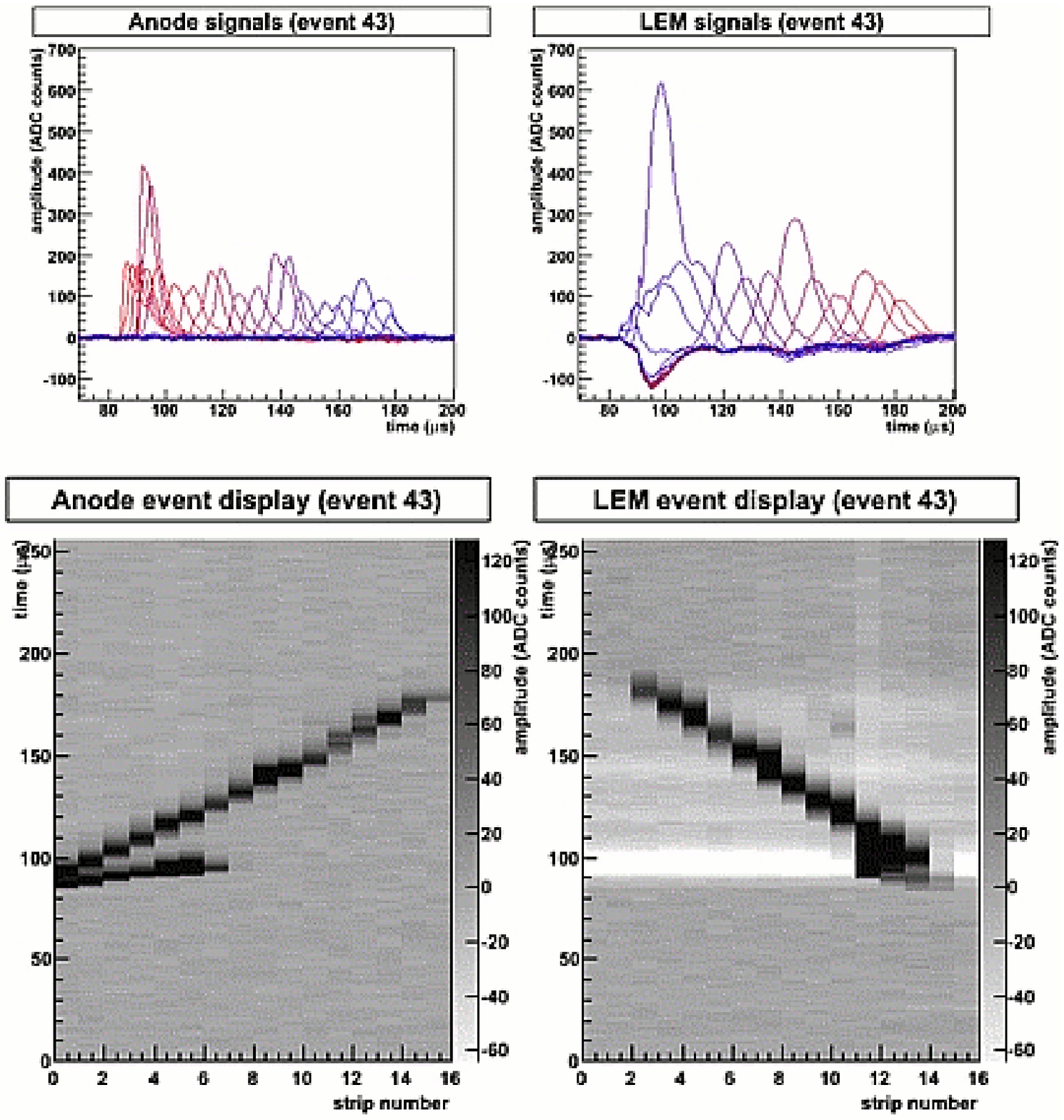}
\end{minipage}}%
\hfill
\subfloat[Landau distribution from crossing muons]{
\label{fig:1mmTHGEM:b} 
\begin{minipage}[b]{0.33\linewidth}
\centering \includegraphics[width=0.85\linewidth]{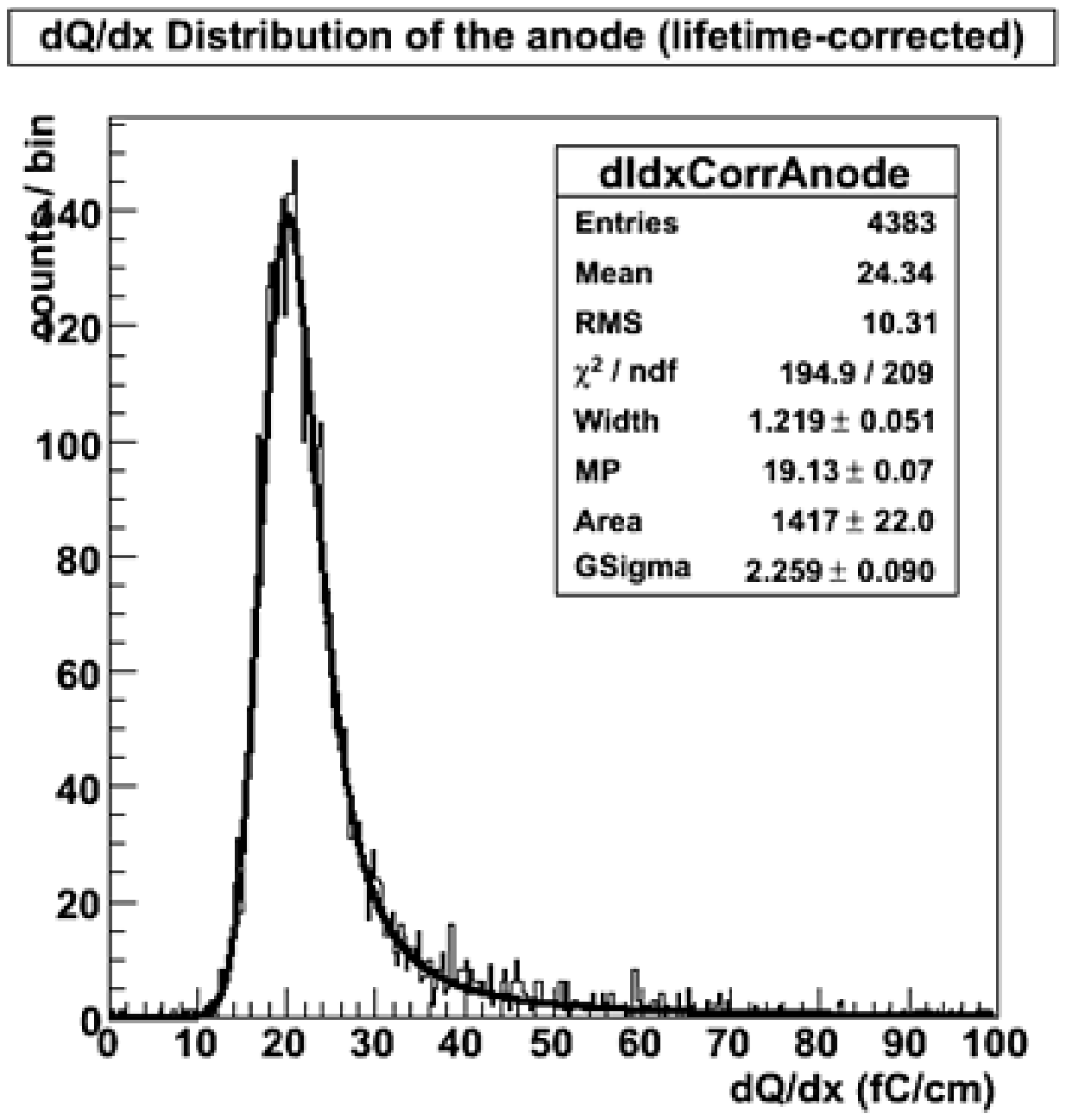}
\end{minipage}}
\caption{Performance of 1 mm thick single stage thick GEM operating at 3.6 kV}
\label{fig:1mm_THGEM} 
\end{figure}

A 3 liters prototype double-phase argon TPC has been successfully operated with single and double stage THGEMs, for the first time with imaging capability, as described in \Refs~\cite{Badertscher:2008rf, Badertscher:2009av}. \Figure[b]~\ref{fig:THGEMTPC} shows the test setup with a double stage THGEM (otherwise called LEM, Large Electron Multiplier) of $10\times 10$ cm$^{2}$. The amplified electron signal is readout via two orthogonal coordinates, the segmented upper electrode of the THGEM itself followed by a segmented anode, both with 6 mm wide strips. Half of the charge is usually collected on the THGEM electrode, while the remaining half reaches the anode. This novel device, which will be referred to as LAr THGEM-TPC, has been proposed in \Ref~\cite{Rubbia:2004tz} for large size LAr detectors.
Examples of cosmic muons are shown in \Fref{fig:1mm_THGEM:a} for a single stage 1 mm thick THGEM. A very good signal/noise ratio of $\sim 60$ is apparent from the digitized signals recorded by the THGEM and anode 6 mm wide strips. This configuration provides, for a potential difference of 3.6 kV across the THGEM, an effective charge gain of $\sim$2-3 for each of the two readout electrodes. The distribution of released charge per unit length from reconstructed cosmic muons using the anode signals is shown in \Fref{fig:1mmTHGEM:b} with a superimposed fitted Landau distribution. Larger charge gains of about 10 have been achieved with a double stage THGEM of 1.6 mm thickness ~\cite{Badertscher:2009av}. Installation of a $\sim$0.5 m$^{2}$ THGEM in the ArDM experiment ~\cite{Rubbia:2005ge} is foreseen in the next few months. The process of industrialization for the production of many squared meters of THGEMs detectors, necessary in large size LAr THGEM-TPCs as well as other applications, is being pursued by the RD51 Collaboration at CERN ~\cite{RD51}.   

The basic design of preamplifiers with multiple low noise JFETs at the input stage, connected in parallel to match a high detector capacitance, has been first employed by the ICARUS collaboration for the charge readout of a LAr TPC ~\cite{Centro:2009} and successfully adopted by other groups. The measurements shown in \Fref{fig:1mm_THGEM} make use of a custom hybrid preamplifier ~\cite{Badertscher:2008rf} with a signal to noise ratio of 10 for 1 fC input charge and 200 pF input capacitance. While this performance is the state of the art for this technology, such a design is not directly transferable to an ASIC design, with a consequent cost reduction, because of the use of discrete JFET components. Attempts to reach similar noise performances by making use of ASIC preamplifiers based on CMOS technology, operated at cryogenic temperatures inside the detector vessel, are currently underway (see \Refs~\cite{Fleming:2009, Girerd:2009}. This solution also allows to minimize the length, and consequently the capacitance, of the cables between the readout electrodes and the preamplifiers. This is particularly important in some of the designs of large size LAr TPCs, because it provides freedom in the location of the readout electrodes. Low power digitizers and multiplexers inside the cryogenic vessel have been proposed in ~\cite{Fleming:2009} as a way to largely reduce the number of feedthroughs.

Since argon is a very good scintillator, recently attempts have been made to detect the released ionization charge in LAr by a radioactive source through the secondary scintillation light produced when ionization electrons are driven into the holes of a THGEM, where a high electric field is present ~\cite{Lightfoot:2008ig}. Signals from an Fe$^{55}$ source have been observed with good resolution in a double-phase argon setup: the scintillation light produced by the electrons extracted from the liquid phase and drifted into the holes of 1.5 mm thick THGEM, operating between 2.1 kV and 3.4 kV, has been detected by a single 1 mm$^{2}$ silicon photomultiplier, coated with a wavelength shifter (tetraphenyl butadiene). Extrapolation of this technique to large size LAr TPCs with tracking resolutions of a few mm  would require an array of photosensors mounted behind the THGEM plane, calling for substantial R\&D on low cost photosensors. A first attempt to produce secondary scintillation light in the holes of the THGEM directly immersed in liquid argon requires more work to fully comprehend the preliminary observations. Signals from the Fe$^{55}$ source have been observed, but for a very narrow range of the applied high voltage across the THGEM ($\sim$10 kV) and with considerable worsening in resolution.
\subsection{LAr vessels}
\label{sec:vessels}
Traditionally high purity LAr vessels have been built as stainless steel vacuum insulated dewars, where the inner vessel is also evacuable. Such devices are not scalable to the required dimensions of the envisioned LAr detectors (> 10000 m$^{3}$), if built with standard techniques. A design trying to maintain the features of vacuum insulation and evacuable vessels for very large containers, with regularly spaced mechanical reinforcements, is described in Section~\ref{sec:designs}. 

The ICARUS T600 vessel, which was designed with the main criteria to be transportable on the Italian highways, utilizes an aluminum LAr vessel with evacuated, honeycomb-structured insulation panels.

Very large cryogenic vessels, with a more favorable volume/surface ratio, can make use of passive insulation. Industrial tanks for the containment of liquefied natural gas (LNG, > 95\% CH$_{4}$), up to volumes of 200000 m$^{3}$, are built as non-evacuable nickel steel tanks, making use of perlite or foam glass as thermal insulation. Several groups have proposed to use similar vessels for the containment of LAr (see \Refs~\cite{mulholland:2002}, ~\cite{Technodyne:2004}, ~\cite{Rubbia:2004tz}, ~\cite{Bartoszek:2004si}) based on the following facts:
the boiling points of LAr and CH$_{4}$ are quite close, 87.3 and 111.6 K respectively;
the latent heat of vaporization per unit volume is the same for both liquids within 5\%;
with a LAr density  3.3 times higher than liquefied CH$_{4}$, the tank needs to withstand 3.3 times higher hydrostatic pressure, which is achievable with thicker steel;
thermal losses of $\sim$5 W/m$^{2}$ are achievable with passive insulation methods ($\sim1.5$ m thick perlite insulation), resulting in a boil-off of only 0.04\%/day for a 100 kton LAr vessel.

A corrugated stainless steel/Invar membrane tank, as recently built for an underground pilot plant for containment of LN$_{2}$ ~\cite{Geostock:2009}, is also being considered ~\cite{Fleming:2009}.

In such designs the LAr vessel is non-evacuable and it requires purification from air before proceeding with the LAr filling. Preliminary tests on small tanks have already been conducted at FNAL ~\cite{Jaskierny:2006sr} and at KEK ~\cite{Maruyama:2009} with gas argon purging, reaching O$_{2}$ contaminations less than 50 ppm within a few hours of purging. Large scale tests are foreseen at CERN by the ETH Zurich group with a 6 m$^{3}$ device, and at FNAL with a 20 ton LAr tank (LAPD project). The purification procedure starting from air will also provide a check of the tightness of the non-evacuable vessels through the measurement of the residual O$_{2}$ and N$_{2}$ contaminations in the argon gas. 
\subsection{High voltage systems}
\label{sec:HV}
A drift field of 0.5 kV/cm, the nominal value in the ICARUS T300 detector over a drift distance of 1.5 m, corresponds to a drift velocity of the ionization electrons of 1.6 mm/$\mu$s. In such a range the drift velocity is already not linear with the electric field, still drift velocities of 2 mm/$\mu$s could be achieved with drift fields of 1 kV/cm, allowing to collect a higher percentage of ionization charge for a given electron lifetime.
\begin{figure}[ht]
\centering
\begin{minipage}[t]{0.25\linewidth}
\centering\includegraphics[width=0.56\linewidth]{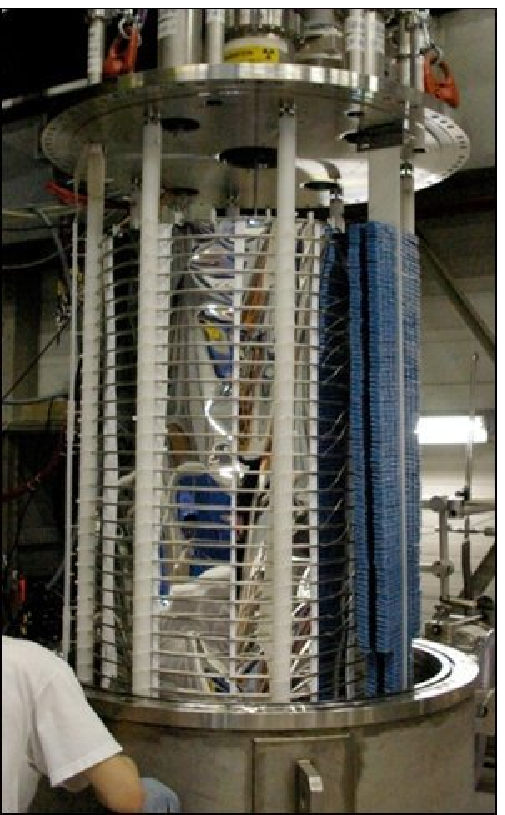}
\caption{Voltage multiplier in the ArDM experiment}\label{fig:ArDM_CockroftWalton}
\end{minipage}%
\hspace{0.1cm}%
\begin{minipage}[t]{0.35\linewidth}
\centering\includegraphics[width=0.9\linewidth]{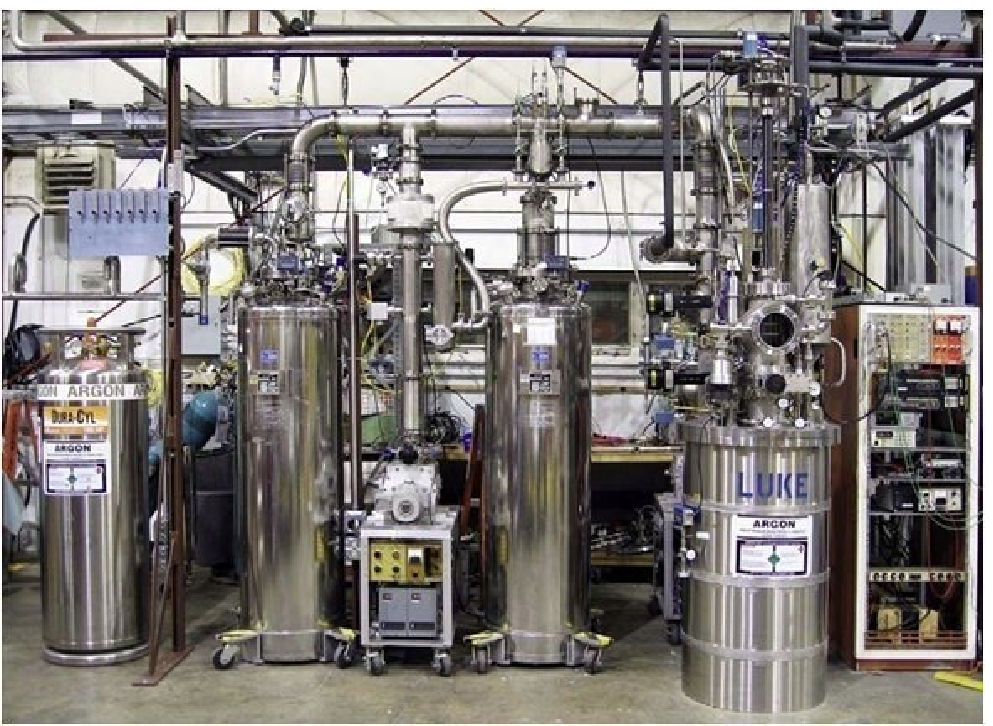}
\caption{Material test stand at FNAL}\label{fig:Material_test_stand_FNAL}
\end{minipage}%
\hspace{0.5cm}%
\begin{minipage}[t]{0.35\linewidth}
\centering\includegraphics[width=.9\linewidth]{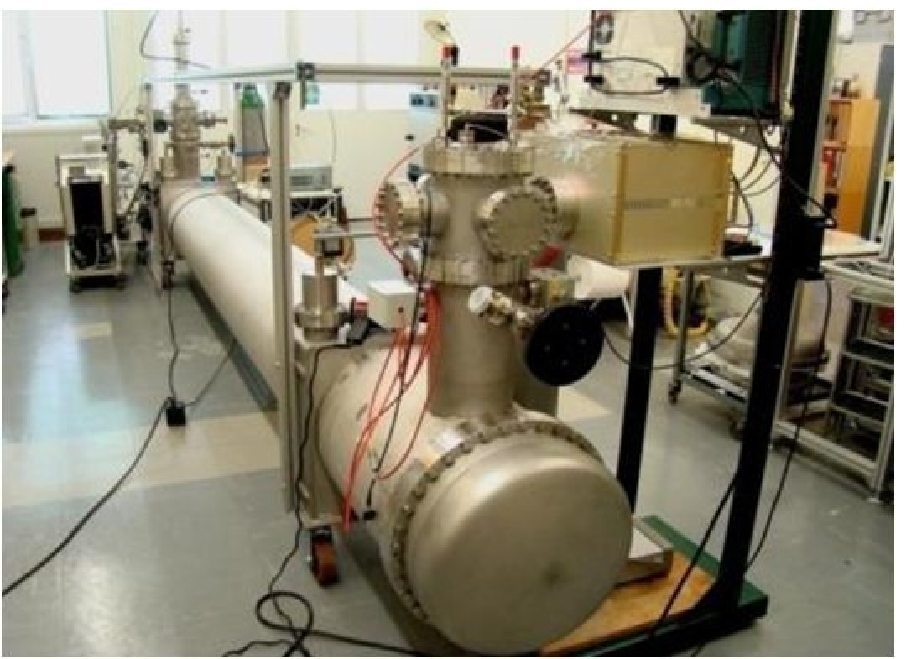}
\caption{Horizontal 5 m drift test stup at CERN}\label{fig:FS_5m}
\end{minipage}
\end{figure}

The need for longer drift paths and relatively high drift velocities drives towards higher high voltage values. At least few 100 kV are envisioned for large LAr detectors. The ICARUS HV feedthrough ~\cite{Amerio:2004ze} has been tested up to 150 kV, and the design could in principle be extrapolated to larger values. A different approach has been followed in the ArDM experiment ~\cite{Rubbia:2005ge, Kaufmann:2007zz}, where a 210 stage Cockroft-Walton voltage multiplier, directly immersed in LAr, with a maximum voltage of $\sim$2 kV/stage and providing a drift field of $\sim$4 kV/cm, has been built as shown in \Fref{fig:ArDM_CockroftWalton}, and will be tested in the next few months. Smaller prototypes have already reached voltages of $\sim$120 kV during short term tests in LAr. This technique could be used to generate higher voltages (MV like), in a range where commercial HV power supplies are not available, to produce drift fields of 1 kV/cm over a $\sim$10 m distance, posing less stringent requirements than in the ArDM case on the voltage multiplier.
\subsection{Argon purity}
\label{sec:purity}
The relatively low drift velocity in LAr, combined with drift paths of at least a few meters, implies drift times $\geq$ few ms. During such long drifts it is essential to ensure the collection of a high percentage of the ionization charge, the electron lifetime $\tau$ being directly related to the O$_{2}$-equivalent impurity concentration $\rho$ by $\tau(\mu s)=300/\rho[ppb]$. The concentration of electronegative impurities must be kept at the level of a few tens of ppt to reach electron lifetimes of the order of 10 ms. Since the fundamental achievements of the ICARUS collaboration, argon purity remains an important R\&D subject, mainly focusing on LAr purification techniques, qualification of materials and monitors of argon purity.

The LAr recirculation system of a $\sim$100 kton scale LAr detector will need many cryogenic pumps working in parallel, with small thermal losses and without compromising the purity of the argon. Every small LAr setup offers the opportunity to test new schemes and new devices for LAr recirculation systems and cryogenic pumps, as for example in the ArDM experiment. Custom made purification cartridges have been developed for ArDM, based on CuO powder, and at FNAL, using copper-coated alumina granules ~\cite{Curioni:2009rt}. Such cartridges are easily regenerable at about $250 ^{\circ}$C in a stream of Ar/H$_{2}$ gas. At FNAL electron lifetimes of $\sim$10 ms, at the upper limit of the instrument range, have been routinely obtained with such cartridges. 

In a 120 liters LAr TPC test facility at the INFN-LNL laboratory ~\cite{Baibussinov:2009gs}, making use of a commercial Oxysorb/Hydrosorb filter as in the ICARUS T300 detector, an eletron lifetime of $(21.4^{+7.3}_{-4.3})$ ms has been achieved for several weeks. Both ~\cite{Badertscher:2009av} and ~\cite{Baibussinov:2009gs} stress the importance of purifying or removing the argon gas during the initial phase of the detector cooling, because it is possibly contaminated by outgassing.     

At FNAL a material test facility has been developed to test a number of materials commonly used in detector construction ~\cite{Andrews:2009zza} (see \Fref{fig:Material_test_stand_FNAL}). No effect on the electron lifetime was measured when the materials were immersed in LAr, but, when positioned in the warmer region of the vapor phase (at $\sim 200$ K), a strong decrease of the electron lifetime was observed, correlated with an increase of water concentration in the vapor phase. It is inferred that water is responsible for the decrease of the lifetime and that water concentrations in the liquid phase at the level of 10 ppt affect the electron lifetime.

A novel monitoring and calibration system, exploiting UV laser ionization in LAr, has been developed by the University of Bern ~\cite{Rossi:2009im}. A pulsed ultraviolet Nd-YAG laser, working on the 4th harmonic
with a wavelength of 266 nm corresponding to 4.66 eV photons, produces tracks in LAr with a signal-to-noise ratio of 80 for 20 mJ laser beam energy. This device could be used for electron lifetime determination, but also for calibration and monitoring of large LAr masses. Straight tracks, laser generated, would provide a measurement of the uniformity of the drift field and the drift velocity.
\subsection{Long drifts}
Electron drifts of at least a few meters are necessary for the realization of a realistic 100 kton scale LAr detector, even if built in a modular way. Full scale measurements of long drifts are necessary to assess the effects of signal attenuation and charge diffusion, and at the same time they represent a test of the high voltage systems and of the capability to achieve and maintain the necessary LAr purity. 

An horizontal drift setup of 5 m has been fully assembled at CERN (see \Fref{fig:FS_5m} ~\cite{Sergiampietri:2009}), where it is being commissioned, while a vertical drift setup, again of 5 m, is under construction at the University of Bern. A more ambitious 10 m long vertical drift test is presently being contemplated at KEK, in a $\sim$30 m$^{3}$ vessel.
\subsection{Large magnetized LAr volumes}
\label{magnet}
Traditionally, bubble chambers and calorimetric neutrino detectors have been magnetized, in order to get a measurement of at least the muon momentum. These detectors also allow, by measuring the muon charge, the determination of the minority antineutrino (neutrino) component in wide band neutrino (antineutrino) beams, and the identification of charmed particles through their semileptonic decays. The presence of a magnetic field is essential in case of a neutrino beam from a neutrino factory for the identification of the so-called right and wrong sign leptons. Water Cerenkov detectors are not compatible with being operated in a magnetic field, making impossible, for example, the separate measurement of neutrino and antineutrino components of atmospheric neutrinos in such detectors.

The excellent tracking and calorimetric capabilities of a LAr TPC allow the measurement of particle energies for contained events. Even in case of through-going particles, their momenta can de determined up to a few GeV by means of multiple scattering. The possibility to magnetize large LAr TPCs has been discussed in \Refs~\cite{Rubbia:2001pk, Cline:2001pt} mainly motivated by the measurement of the sign of the leptons. While relatively low fields (B$\gtrsim$0.1 T) are sufficient to determine the charge of muon tracks at least a few meters long, higher fields (B$\gtrsim$1 T) are necessary for the determination of the charge of electrons of a few GeV, since they start showering quite early in LAr (14 cm radiation length) ~\cite{Rubbia:2001pk}. A first proposal in \Ref~\cite{ Cline:2001pt} considered a huge vertical warm solenoid enclosing the LAr volume, surrounded by an iron iron yoke for the flux return. The warm coil would dissipate 17 MW for a field B=0.2 T, raising questions for the thermal insulation of the LAr vessel. To avoid this problem \Ref~\cite{Ereditato:2005yx} proposes to immerse a superconducting solenoid directly into the LAr vessel, possibly built out of High T$_{c}$ superconductor cable, which could be operated at a temperature substantially larger than 4 K and possibly at the LAr temperature. Since HTS cables find many technological applications, e.g. Superconducting Magnetic Energy Storage, this is a rapidly developing market which needs to be carefully monitored.

A parameter directly connected with the complexity of a magnet is its total stored energy. A 100 kton LAr detector, magnetized at the maximum contemplated field of B=1 T, would have a stored magnetic energy of 30 GJ, about a factor ten higher than the energy stored in the solenoid of the CMS experiment at LHC. This is an extrapolation that might not be unrealistic, but that certainly requires dedicated engineering studies.

\section{Summary of the main design concepts}
\label{sec:designs}
Motivated by the necessity of large detectors for neutrino physics, proton decay and the observation of astrophysical neutrinos, several designs of large LAr detectors, up to $\sim$100 kton scale, have appeared in the last several years (see \Refs~\cite{Rubbia:2004tz, Rubbia:2009md},~\cite{Bartoszek:2004si},~\cite{Cline:2001pt, Cline:2006st},~\cite{Baibussinov:2007ea, Angeli:2009zza},~\cite{Fleming:2009}). 
\begin{figure}[ht]
\centering
\begin{minipage}[t]{0.27\linewidth}
\centering\includegraphics[width=1.0\linewidth]{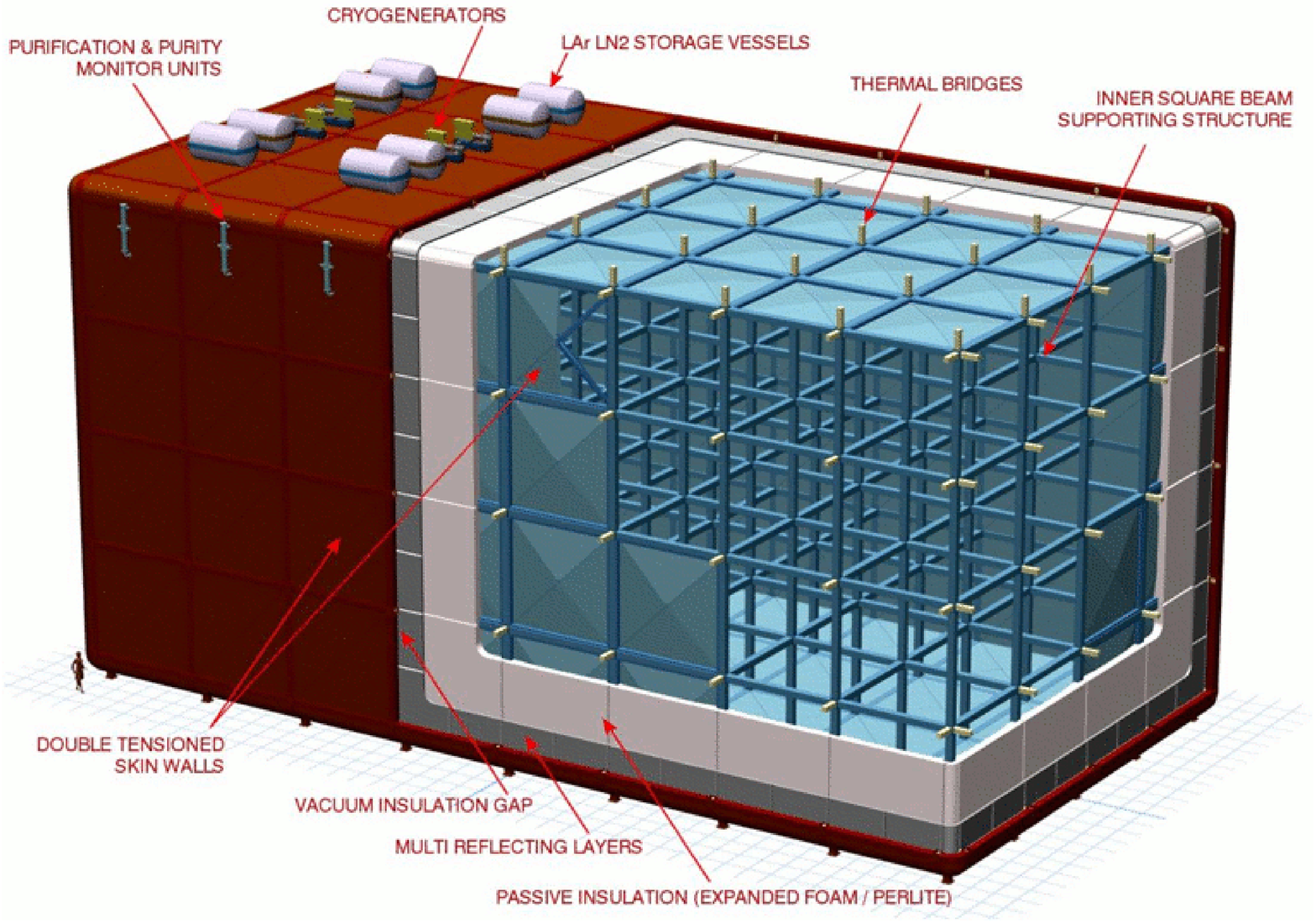}
\caption{Evacuable cryostat concept}\label{fig:LANDD_Cryostat}
\end{minipage}
\hspace{0.2cm}
\begin{minipage}[t]{0.31\linewidth}
\centering\includegraphics[width=1.0\linewidth]{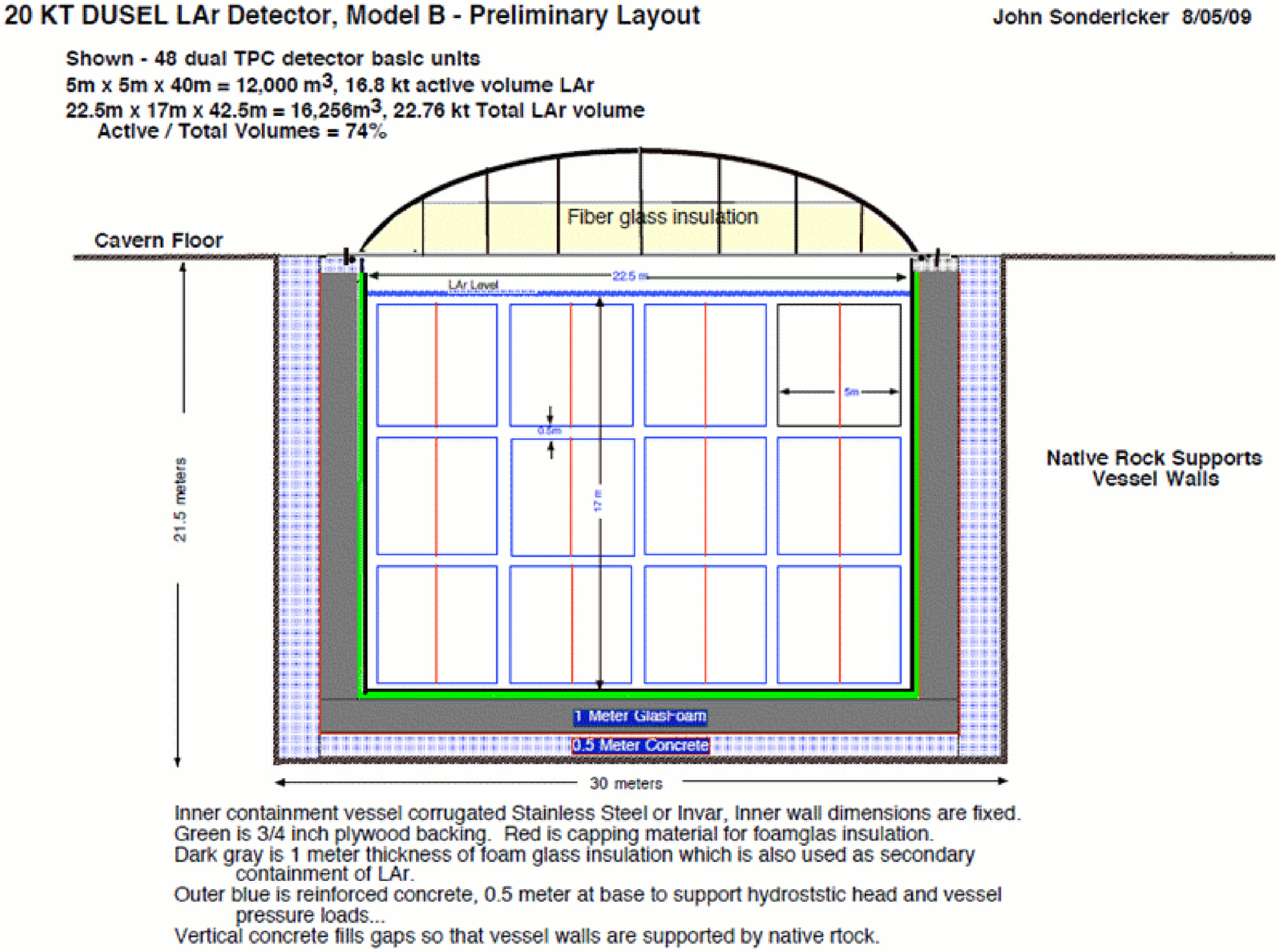}
\caption{LBNE cryostat concept}\label{fig:LBNE_Cryostat}
\end{minipage}
\begin{minipage}[t]{0.39\linewidth}
\centering\includegraphics[width=1.0\linewidth]{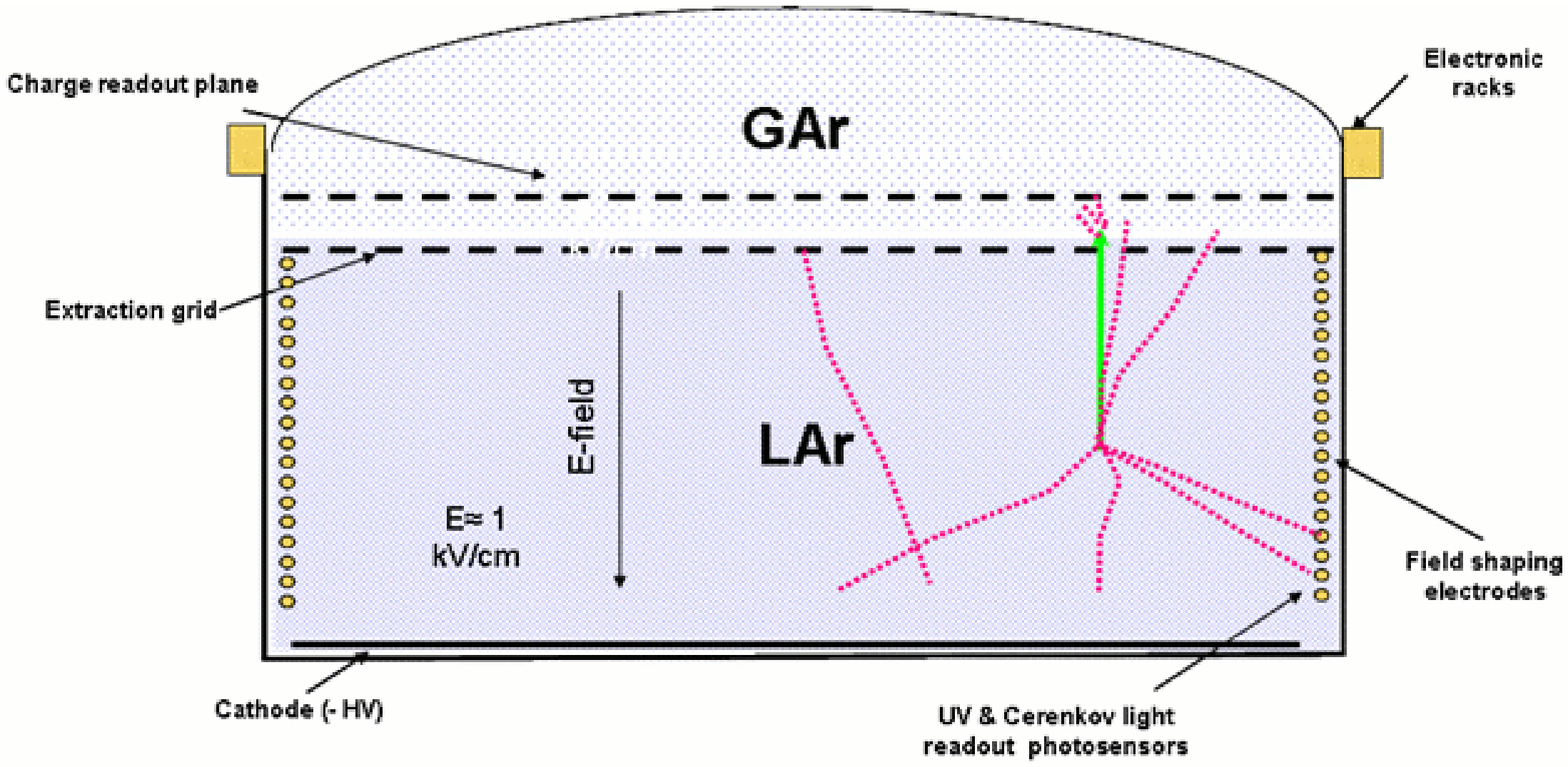}
\caption{GLACIER design concept}\label{fig:Glacier}
\end{minipage}
\end{figure}

\Ref[b]~\cite{Cline:2006st} proposes scalable LAr TPC detectors based on an array of 3-dimensional cube frames immersed in a common LAr volume (see \Fref{fig:LANDD_Cryostat}). This mechanical structure, if on one side complicates the construction of the readout devices, has been designed to allow evacuation of the inner vessel and consequently to preventively check its tightness. The design assumes a double wall cryostat, with vacuum insulation around the cold vessel. 

Within the LBNE project in the US, it is being developed a conceptual design for an initial LAr module of 20 kton, to be installed in the DUSEL laboratory ~\cite{Fleming:2009}. A total active mass of 50 to 100 kton would be achieved by the construction of multiple 20 kton modules. \Figure[b]~\ref{fig:LBNE_Cryostat} shows one design concept, with large rectangular TPC modules stacked inside a corrugated stainless steel/Invar membrane tank, using the cavern walls as support. 

The MicroBooNE experiment at FNAL ~\cite{Chen:2007zz}, with 100 ton LAr active volume, is considered an important step in the development of LAr TPCs in the US. In particular MicroBooNE will test the achievable purity when filling the vessel with LAr without evacuation, by initially purging with  gaseous argon. The MicroBoone collaboration will implement hybrid JFET preampliers working in cold argon gas, investigating at the same time the development of cold CMOS electronics in LAr.

A single LAr module of 100 kton, denominated GLACIER, has been proposed in \Ref~\cite{Rubbia:2004tz, Rubbia:2009md} (see \Fref{fig:Glacier}), with an almost total active LAr mass. The vessel is a cylindrical cyrogenic tank based on industrial LNG technology, providing an excellent volume/surface ratio, thus minimizing thermal losses and the effect of outgassing from the container walls. In order to reach 100 kton LAr mass, a single cylinder of 70 m diameter and 20 m height is being considered, with a single long vertical drift at high drift field ($\sim$1 kV/cm), provided by an immersed Cockroft-Walton voltage multiplier. With a corresponding electron drift velocity of 2 mm/$\mu$s, this would allow better than 30\% charge collection over 20 m drift for an electron lifetime of 10 ms. The charge loss during such long drifts can be compensated by operating the device in double-phase argon and by using charge amplification in pure argon gas by THGEMs, as described in Section ~\ref{sec:readout}. 

As part of the R\&D path to successfully design and propose a GLACIER-like detector, in addition to the R\&D projects already described, it is proposed to expose a LAr TPC detector of about 6 m$^{3}$, operated in double-phase, to test beams in CERN SPS North Area ~\cite{6m3} and to construct an engineering prototype of 1 kton ~\cite{Rubbia:2009md}, that could provide important physics outputs.
A 1 kton detector, realized as a cylinder of 12 m diameter and 10 m height, is the largest possible detector that minimizes construction timescale and costs, as well as the extrapolation to 100 kton. All tank and LAr purification issues would be addressed by such a device and only a factor two extrapolation in drift length would be required for the final 100 kton tank. Underground construction and operation of a GLACIER-like tank, including all safety issues, is being studied as part of the LAGUNA design study ~\cite{LAGUNA:2009} for specific european sites.
\section{Conclusions}
A 100 kton scale LAr detector aims at physics with discovery potential: the mass ordering of neutrinos, CP violation in the leptonic sector, proton decay, and more generally addressing fundamental physics at the Grand Unified scale. Such a detector, if built in Europe, would clearly benefit from a neutrino beam from CERN. A large choice of underground facilities at different baselines is being studied in the LAGUNA project. 

Several proposals for large LAr TPCs have been put forward in different regions of the world, following somewhat different design concepts, but with many basic R\&D issues in common. R\&D for LAr, first pioneered by the ICARUS collaboration at CERN, is now being actively pursued in Europe, US and Japan. The ICARUS T600 detector, representing the first large scale LAr underground installation, is about to be commissioned and the ArgoNeut chamber on the NuMI neutrino beam at FNAL is recording neutrino interactions.

Some noteworthy R\&D results for large scale detectors have been achieved in these last years: a 3 liters LAr THGEM-TPC prototype operated in double-phase has been shown to be able to achieve a better signal/noise ratio, new methods and technologies to purify LAr have resulted in lifetimes in excess of 10 ms, novel devices to monitor argon purity and drift velocities over large LAr masses have been developed. R\&D on cold electronics, with far reaching consequences, is just started.

While engineering studies on the assembly and operation of 100 kton scale LAr vessels in underground locations are proceeding, several smaller scale setups will start operation in the next few years, for tests of novel high voltage systems, large scale operation of double-phase argon TPCs, full scale measurements of long drifts in LAr, large scale test of argon purification in non-evacuable vessels. The results from these setups will be essential to proceed with a reliable proposal and a sound cost estimate for a 100 kton scale LAr detector. Such a detector will probably require the construction of a full engineering prototype of 1 kton scale, which by itself would provide interesting physics output if located on a short-baseline neutrino beam.
\section*{Acknowledgements}
I warmly acknowledge useful discussions with B. Baller, A. Rubbia and F. Sergiampietri and the many interactions with my colleagues from ETH Zurich and the LAr KEK group.


\begin{thebibliography} {99}
\bibitem{Rubbia:2004tz}
  A.~Rubbia,
  \emph{Experiments for CP-violation: A giant liquid argon scintillation,  Cerenkov
  and charge imaging experiment?},
  arXiv:hep-ph/0402110.
\bibitem{Bartoszek:2004si}
  L.~Bartoszek {\it et al.},
  \emph{FLARE: Fermilab liquid argon experiments},
  arXiv:hep-ex/0408121.
\bibitem{Meregaglia:2006du}
  A.~Meregaglia and A.~Rubbia,
  \emph{Neutrino oscillation physics at an upgraded CNGS with large next
  generation liquid argon TPC detectors},
  JHEP {\bf 0611} (2006) 032
  [arXiv:hep-ph/0609106].
\bibitem{Barger:2007yw}
  V.~Barger {\it et al.},
  \emph{Report of the US long baseline neutrino experiment study},
  arXiv:0705.4396 [hep-ph].
\bibitem{Baibussinov:2007ea}
  B.~Baibussinov {\it et al.},
  \emph{A new, very massive modular Liquid Argon Imaging Chamber to detect low
  energy off-axis neutrinos from the CNGS beam. (Project MODULAr)},
  Astropart.\ Phys.\  {\bf 29} (2008) 174
  [arXiv:0704.1422 [hep-ph]].
\bibitem{Badertscher:2008bp}
  A.~Badertscher {\it et al.},
  \emph{A Possible Future Long Baseline Neutrino and Nucleon Decay Experiment with a 100 kton Liquid Argon TPC at Okinoshima using the J-PARC Neutrino
  Facility},
  arXiv:0804.2111 [hep-ph].
\bibitem{Bueno:2007um}
  A.~Bueno {\it et al.},
  \emph{Nucleon decay searches with large liquid argon TPC detectors at shallow
  depths: Atmospheric neutrinos and cosmogenic backgrounds},
  JHEP {\bf 0704} (2007) 041
  [arXiv:hep-ph/0701101].
\bibitem{GilBotella:2004bv}
  I.~Gil Botella and A.~Rubbia,
  \emph{Decoupling supernova and neutrino oscillation physics with LAr TPC
  detectors},
  JCAP {\bf 0408} (2004) 001
  [arXiv:hep-ph/0404151].
\bibitem{Cocco:2004ac}
  A.~G.~Cocco, A.~Ereditato, G.~Fiorillo, G.~Mangano and V.~Pettorino,
  \emph{Supernova relic neutrinos in liquid argon detectors},
  JCAP {\bf 0412} (2004) 002
  [arXiv:hep-ph/0408031].
\bibitem{CRubbia:1977}
C.~Rubbia, \emph{The Liquid-Argon Time Projection Chamber:A New Concept For Neutrino Detector}, CERN-EP/77-08(1977).
\bibitem{Amerio:2004ze}
  S.~Amerio {\it et al.}  [ICARUS Collaboration],
  \emph{Design, construction and tests of the ICARUS T600 detector},
  Nucl.\ Instrum.\ Meth.\  A {\bf 527} (2004) 329.
\bibitem{Arneodo:2006ug}
  F.~Arneodo {\it et al.}  [ICARUS-Milano Collaboration],
  \emph{Performance of a liquid argon time projection chamber exposed to the  WANF neutrino beam},
  Phys.\ Rev.\  D {\bf 74} (2006) 112001
  [arXiv:physics/0609205].
\bibitem{Fleming:2009}
 B.~Fleming, \emph{LAr detector R\&D in the US}, talk at NNN09, Estes Park, Oct. 2009.
\bibitem{Maruyama:2009}
 T.~Maruyama, \emph{LAr detector R\&D in Japan}, talk at NNN09, Estes Park, Oct. 2009.
\bibitem{Gatti:1979}
 E.~Gatti {\it et al.},
  \emph{Considerations for the design of a time projection liquid argon ionization chamber}, IEEE Trans. Nucl. Sci. {\bf 26} (1979) 2910.
\bibitem{Angeli:2009zza}
  D.~Angeli {\it et al.},
  \emph{Towards a new Liquid Argon Imaging Chamber for the MODULAr project},
  JINST {\bf 4} (2009) P02003.
\bibitem{Breskin:2008cb}
  A.~Breskin {\it et al.},
  \emph{A concise review on THGEM detectors},
  Nucl.\ Instrum.\ Meth.\  A {\bf 598} (2009) 107
  [arXiv:0807.2026 [physics.ins-det]].
\bibitem{Alon:2008zz}
  R.~Alon {\it et al.},
  \emph{Operation of a thick gas electron multiplier (THGEM) in Ar, Xe and Ar-Xe},
  JINST {\bf 3} (2008) P01005.
\bibitem{Polina_ETHthesis}
  P.~Otyugova,
  \emph{Development of a Large Electron Multiplier~(LEM) based charge readout system for the ArDM experiment}, Diss. ETH No. 17704, 2008.
\bibitem{Bondar:2008yw}
  A.~Bondar, A.~Buzulutskov, A.~Grebenuk, D.~Pavlyuchenko, Y.~Tikhonov and A.~Breskin,
  \emph{Thick GEM versus thin GEM in two-phase argon avalanche detectors},
  JINST {\bf 3} (2008) P07001
  [arXiv:0805.2018 [physics.ins-det]].
\bibitem{RD51}
RD51 Collaboration, \url{http://rd51-public.web.cern.ch/RD51-Public/Welcome.html}
\bibitem{Badertscher:2008rf}
  A.~Badertscher {\it et al.},
  \emph{Construction and operation of a Double Phase LAr Large Electron Multiplier Time Projection Chamber},
  arXiv:0811.3384 [physics.ins-det].
\bibitem{Badertscher:2009av}
  A.~Badertscher {\it et al.},
  \emph{Operation of a double-phase pure argon Large Electron Multiplier Time
  Projection Chamber: comparison of single and double phase operation},
  Nucl.Instr.andMeth.A(2009), doi:10.1016/j.nima.2009.10.011
[arXiv:0907.2944 [physics.ins-det]].
\bibitem{Rubbia:2005ge}
  A.~Rubbia,
  \emph{ArDM: A ton-scale liquid argon experiment for direct detection of dark
  matter in the universe},
  J.\ Phys.\ Conf.\ Ser.\  {\bf 39} (2006) 129
  [arXiv:hep-ph/0510320].
\bibitem{Centro:2009}
S.~Centro, \emph{Cost effective electronics for LAr and photodetectors readout}, to appear in CERN Yellow Report - Future Neutrino Physics Workshop, CERN, October 2009
\bibitem{Girerd:2009}
C.~Girerd {\it et al.}, \emph{Cold front-end electronics and Ethernet-based DAQ system for large LAr TPC readout}, to appear in CERN Yellow Report - Future Neutrino Physics Workshop, CERN, October 2009
\bibitem{Lightfoot:2008ig}
  P.~K.~Lightfoot, G.~J.~Barker, K.~Mavrokoridis, Y.~A.~Ramachers and N.~J.~C.~Spooner,
  \emph{Optical readout tracking detector concept using secondary scintillation
  from liquid argon generated by a thick gas electron multiplier},
  JINST {\bf 4} (2009) P04002
  [arXiv:0812.2123 [physics.ins-det]].
\bibitem{mulholland:2002}
G.T.~Mulholland, \emph{LANNDD Feasibility Study}, Aug. 13, 2002, \url{http://www.hep.princeton.edu/~mcdonald/nufact/mulholland/ELAN_Proposal.pdf}
\bibitem{Technodyne:2004}
Technodyne International Ltd, \emph{Large Underground Argon Storage Tank
Study report}, December 2004, commissioned by ETH Zurich.
\bibitem{Geostock:2009}
G\'eostock Group, \url{http://www.geostockgroup.com}
\bibitem{Jaskierny:2006sr}
  W.~Jaskierny, H.~Jostlein, S.~H.~Pordes, P.~A.~Rapidis and T.~Tope,
  \emph{Test of purging a small tank with argon}, FERMILAB-TM-2384-E.
\bibitem{Kaufmann:2007zz}
  L.~Kaufmann and A.~Rubbia,
  \emph{The ArDM Project: A Direct Detection Experiment, Based On Liquid Argon, For The Search Of Dark Matter},
  Nucl.\ Phys.\ Proc.\ Suppl.\  {\bf 173} (2007) 141.
\bibitem{Curioni:2009rt}
  A.~Curioni {\it et al.},
  \emph{A Regenerable Filter for Liquid Argon Purification},
  Nucl.\ Instrum.\ Meth.\  A {\bf 605} (2009) 306
  [arXiv:0903.2066 [physics.ins-det]].
\bibitem{Baibussinov:2009gs}
  B.~Baibussinov {\it et al.},
  \emph{Free electron lifetime achievements in Liquid Argon Imaging TPC},
  arXiv:0910.5087 [physics.ins-det].
\bibitem{Andrews:2009zza}
  R.~Andrews, W.~Jaskierny, H.~Jostlein, C.~Kendziora, S.~Pordes and T.~Tope,
  \emph{A System To Test The Effects Of Materials On The Electron Drift Lifetime In Liquid Argon And Observations On The Effect Of Water},
  Nucl.\ Instrum.\ Meth.\  A {\bf 608} (2009) 251.
\bibitem{Rossi:2009im}
  B.~Rossi {\it et al.},
  \emph{A prototype liquid Argon Time Projection Chamber for the study of UV laser multi-photonic ionization},
  JINST {\bf 4} (2009) P07011
  [arXiv:0906.3437 [physics.ins-det]].
\bibitem{Sergiampietri:2009}
Figure courtesy of F. Sergiampietri.
\bibitem{Rubbia:2001pk}
  A.~Rubbia,
  \emph{Neutrino factories: Detector concepts for studies of CP and T violation
  effects in neutrino oscillations},
  arXiv:hep-ph/0106088.
\bibitem{Cline:2001pt}
  D.~B.~Cline, F.~Sergiampietri, J.~G.~Learned and K.~McDonald,
  \emph{LANNDD: A massive liquid argon detector for proton decay, supernova and
  solar neutrino studies, and a neutrino factory detector},
  Nucl.\ Instrum.\ Meth.\  A {\bf 503} (2003) 136
  [arXiv:astro-ph/0105442].
\bibitem{Ereditato:2005yx}
  A.~Ereditato and A.~Rubbia,
  \emph{Conceptual design of a scalable multi-kton superconducting magnetized
  liquid argon TPC},
  Nucl.\ Phys.\ Proc.\ Suppl.\  {\bf 155} (2006) 233
  [arXiv:hep-ph/0510131].
\bibitem{Rubbia:2009md}
  A.~Rubbia,
  \emph{Underground Neutrino Detectors for Particle and Astroparticle Science: the Giant Liquid Argon Charge Imaging ExpeRiment (GLACIER)},
  J.\ Phys.\ Conf.\ Ser.\  {\bf 171} (2009) 012020
  [arXiv:0908.1286 [hep-ph]].
\bibitem{Cline:2006st}
  D.~B.~Cline, F.~Raffaelli and F.~Sergiampietri,
  \emph{LANNDD: A line of liquid argon TPC detectors scalable in mass from
  200-tons to 100-ktons},
  JINST {\bf 1} (2006) T09001
  [arXiv:astro-ph/0604548].
\bibitem{Chen:2007zz}
  H.~Chen {\it et al.}  [MicroBooNE Collaboration],
  \emph{Proposal for a New Experiment Using the Booster and NuMI Neutrino
  Beamlines: MicroBooNE}, FERMILAB-PROPOSAL-0974.
\bibitem{6m3}
D.~Autiero {\it et al.},
\emph{Test beam exposure of a Liquid Argon TPC Detector at the CERN SPS North Area}, Abstract \#82, Workshop on New Opportunities in the Physics Landscape at CERN, May 2009
\bibitem{LAGUNA:2009}
The LAGUNA design study is financed by FP7 Research Infrastructure ``Design Studies'', Grant Agreement No. 212343 FP7-INFRA-2007-1. See \url{http://laguna.ethz.ch}
\end{thebibliography}
\end{document}